# Quantifying structure-property relationships during resistance spot welding of an aluminum 6061-T6 joint

S.A. Turnage[1], K.A. Darling[2], M. Rajagopalan[1], W.R. Whittington[3], M.A. Tschopp[2], P. Peralta[1], K.N. Solanki[1*]

[1]School for Engineering of Matter, Transport, and Energy, Arizona State University, Tempe, AZ 85287
[2]Army Research Laboratory, Weapons and Materials Research Directorate, APG, MD 21014
[3]Center for Advanced Vehicular Systems, Mississippi State University, Starkville, MS 39759

*Corresponding Author: (480) 965-1869; (480) 727-9321 (fax), E-mail: kiran.solanki@asu.edu

## Abstract

Microstructure-property relationships of resistance spot welded 6061-T6 aluminum alloy lap joints were investigated via mechanical testing and microscopy techniques. Quasi-static tensile and novel shear punch tests were employed to measure the mechanical properties of the distinct weld regions. Quasi-static tensile and shear punch tests revealed constantly decreasing strength and ductility as the weld center was approached. For instance, the ultimate tensile strength of the fusion zone decreased by ~52% from the parent material (341 MPa to 162 MPa) while the yield strength decreased by ~62% (312 MPa to 120 MPa). The process-induced microstructures were analyzed with scanning electron microscopy and optical microscopy to elucidate the underlying cause of the reduced mechanical properties. Fractography reveals void growth from particles being the dominant damage mechanism in the parent material as compared to void nucleation in the fusion zone. Overall, significant changes in the mechanical behavior across the weld are the result of a change in microstructure congruent with a loss of T6 condition (precipitate coarsening).

## 1. Introduction

The automotive and defense industries require welding in lightweight alloys [1, 2] to decrease the weight of ground vehicles, thereby addressing energy and emission concerns, improving fuel economy, and reducing production costs. Great advances have been made in understanding weld microstructures especially in steels [3, 4], but design of joints in lightweight alloys is still limited as there are gaps in the understanding of the interaction of lightweight alloy microstructural changes with the resulting change in mechanical behavior as a result of welding processes. Accomplishing this is no small task as it requires new lightweight structural alloys and *joining techniques* that can efficiently form these metals into components. In fact, brittle mechanical responses and other limitations of existing joining techniques (e.g., resistance spot welding [RSW], friction stir welding, etc.) currently hinder the widespread use of lightweight alloys. Attempts have been made to characterize the effects of welding processes on aluminum alloys [5–9]. In particular, RSW is commonly considered for joining lightweight alloys in the automotive and defense sectors. However, one limitation of traditional RSW in ultra-high strength steel, for example, is that brittle weld microstructures are produced, thereby restricting RSW use [1, 10].

As illustrated in Figure 1a, the RSW process joins two metal sheets through compression between a pair of water-cooled electrodes with an external applied force. Low-voltage, high-amperage electric current passes through the sheets for a short duration via the two electrodes, generating concentrated heating at the contact surface. Due to both the heat generation at the contact surface and Joule heating, a molten zone forms at the intersection of the two sheets. After the current flow ceases, the electrode force is maintained for another short duration to allow the work-piece to rapidly cool and solidify. The contact surface area depends on the electrode diameter, applied force, temperature, and metal deformation. These welding operations often involve intense thermal gradients and melt convection that result in a continually evolving microstructure away from that of the parent material. Particularly in RSW, an intense melt flow is caused by the induced electromagnetic field, which leads to heterogeneity and anisotropy within the material microstructure. These residual microstructures, present in crystalline materials after processing, influence the overall strength and performance of the joints. As such, quantifying the influence of process-induced microstructural changes (structure) and subsequent effects on mechanical properties is technologically important in advanced energy, transportation, and manufacturing industries. To this end, several researchers have thoroughly investigated the effect of RSW mechanical properties, e.g., Zhang and Senkara [11] and Williams and Parker [12] have compiled detailed reviews of several experimental and theoretical studies investigating the mechanical and physical properties of RSW joints, but some questions regarding the link between processing effects on microstructure and mechanical strength throughout the weld still remain.

Quantifying the deleterious effects of thermo-mechanical-electrical joining processes (i.e., RSW) on the strength and failure behavior of metal components is a highly complex undertaking. In fact, experimental studies and computational modeling of the RSW processes to understand the effect of weld process parameters on joint strength and failure have received much attention in the last several decades [8, 10–33]. Bowden and Williamson were the first to investigate interface behavior between two contacting solids and showed that surface asperities concentrated the current density within the contact region, which caused temperature to rise at the interface when electric current flow occurred [15]. More recently, Khan et al. [34] varied parameters such as epoxy curing time, welding time, welding pressure, welding current, and surface roughness to determine the effects on ultimate shear strength in a single lap bonded joint. They found that each parameter could be optimized to increase the mechanical strength of the joint. A detailed theoretical and experimental study by Bentley and Greenwood [16] showed that the contact resistance played a major role only in the very early stages of heat generation and became less influential in the later stages of the weld nugget formation. Using the one-dimensional heat models of Bowden and Williamson [15] and Greenwood [16], Gould [18] studied the RSW process by including the electrode geometry, internal heat generation, phase change, temperature-dependent material properties, and contact resistance. Through this, Gould [18] showed that the predicted nugget sizes were much larger than those observed in the experiment and concluded that this discrepancy was evidently due to neglecting the radial heat loss to the surrounding sheet.

More recently, Nied [17], Sheppard [21], Tsai et al. [22, 23], Browne et al. [24, 25], Dong et al. [27], Khan et al. [29, 30], Sun and Dong [31], Wu et al. [35], Afshari et al. [36], and Murakawa et al. [26] all performed theoretical and experimental studies to address different aspects of the RSW process, including nugget formation, electrode design, welding parameters, and electrode wear. Ninshu and Murakawa [32] experimentally and numerically investigated the nugget formation process purely based on nugget size. They concluded minimum weldability conditions are required for a RSW of high strength stainless steel sheets. Hamedi and Pashazadeh [33] used a phenomenological model to study the effect of weld process parameters on nugget formations and concluded that the formed fusion zone is large at low electrode forces and high welding currents.

Deng et al. [37] analyzed the effect of sheet thickness on the measured shear stresses in both symmetric peel and lap shear. They found that stress concentrations were highest along the boundary of the weld nugget. Florea et al. [7, 8, 38] investigated the effect of weld current on RSW quality and found that depth of the weld nugget varies linearly with weld current. They also noted that weld current affected the microstructure, shear strength, and fatigue behavior of the lap joint. Thornton et al. [39] investigated the effects of RSW on the resultant strength of aluminum alloy lap joints using overlap tensile shear and coach peel tests, and showed that weld diameter is the most important parameter with regard to the strength of the weld. They also noted that porosity, material expulsion, and cracking only had a small adverse effect on fatigue resistance of the welds [39]. Afshari et al. [40] and Florea et al. [7] analyzed the residual stresses throughout the weld using finite element analysis and x-ray or neutron diffraction. Both investigations found that residual stresses peaked within the center of the weld then greatly reduced toward the edge of the weld nugget. Despite all of the research regarding RSW, the intense changes in microstructure and the effects on shear properties (including spatial variation across the weld) of the weld joint have received little attention. Furthermore, the effects of changes in local microstructures on weld ductility have not been sufficiently investigated (e.g., dendrite formation and precipitate coarsening).

Understanding shear properties and subsequent structure-property relationships has significant consequences for lightweight materials in emergent technologies, including transportation applications, where lightweight materials are sought to reduce emissions and improve performance and efficiency. Currently the lack of complete understanding of the effects of welding processes on the structure-property relationship of lightweight alloys such as 6061-T6 aluminum and high strength steel alloys hinders the use of these materials in high performance applications. To our knowledge, shear properties as a function of position throughout the entire weld have not been determined either. A systematic characterization of structure-property relationships of the individual sections of welded joints is an important first step in the path to control mechanical properties through the use of design of experiment approaches (e.g., varying weld process parameters).

In this study, we use a novel experimental technique to characterize and develop a structure-property database of RSW 6061-T6 aluminum lap joints for use in finite element analyses. More specifically, the mechanical properties are determined from quasi-static tensile tests along with quasi-static shear punch tests through the weld regions.

## 2. Experimental Procedures

For this study, a servo-gun with weld control and copper-zirconium alloy electrodes was used to manufacture the specimens from a sheet of 6061-T6 aluminum. The aluminum sheets were approximately 127 mm long, 38.1 mm wide, and 2 mm thick. The power supply and current transformer used was a mid-frequency direct current power transformer with an 8 V secondary voltage. Water was applied as a cooling agent at a rate of 4 liters/minute. For further detail about the weld sample preparation please refer to [8]. Electrode force (3.8 kN), weld time (0.115 s), and weld current (30 kA) (see [8]) were manually optimized to produce a minimum nugget size of 5.7 mm with minimum shearing force of 3.8 kN per weld. The optimized welds met or exceeded the MIL-W-6858D military specification [41]. The weld process is outlined in Figure 1a and the weld parameters are listed in Table 1.

Small, flat, dog-bone shaped specimens were cut from the center of the weld nuggets as well as the parent material using wire electric discharge machining (EDM) for quasi-static tensile testing. Specimens were cut with gauge length along the rolling direction of the aluminum sheet. Dog-bone specimens can be seen in Figure 1b. Dimensions for these specimens can be found in Table 2. Quasi-static tensile tests were performed at a strain rate of $10^{-3}$ $s^{-1}$ on an electromechanically driven load frame. No fewer than three tests were run for the weld region and the parent material. Images were taken normal to the fracture surfaces using scanning electron microscopy (SEM) after the testing was completed. These tests revealed a significant reduction in tensile strength and ductility for the weld section.

Shear punch testing (SPT) utilizing a punch diameter of approximately 1 mm with punch-die clearance of approximately 30 μm was performed through the weld in the rolling, transverse, and normal directions of the original rolled plates. Specimens for SPT were prepared by slicing 1 mm thick sections and mechanically grinding the specimens to a 4000 grit surface finish. The fine finish is intended to reduce roughness induced stress concentrations as discussed in Guduru et al. [42]. As can be seen in Figure 2, the SPT fixture drives the solid cylindrical punch through the thin plate specimen into a hollow cylindrical die. Here, the displacement is normalized by the specimen thickness providing a consistent comparison between the test results [42]. Consecutive tests were spaced approximately 1.5 mm apart from the center of each test indent. The comparison of the mechanical behavior at different sections of the weld region shows the profound effect of the solidification-induced microstructure.

Microhardness tests were also performed through the weld along the transverse direction to verify SPT results and further investigate the role of process-induced microstructure on the material strength. Vickers hardness was measured through the parent material, HAZ, and fusion zone at points spaced 0.25 mm (0.01 inches) apart. Tensile yield and ultimate strengths correspond to the microhardness by a constant factor that varies as a result of strain hardening in the material [43]. Variability in the strain hardening behavior resulting from the welding process causes the correlation between hardness and tensile strengths to vary throughout the weld. A best estimate of the correlation factor was determined by averaging the appropriate factors for the parent material and fusion zone. The factors are defined as:

$$c_{Yield} = \frac{\sigma_{Yield}}{VH} \tag{1}$$

$$c_{Ultimate} = \frac{\sigma_{Ultimate}}{VH} \tag{2}$$

where $\sigma_{Yield}$ is the tensile yield strength, $\sigma_{Ultimate}$ the ultimate tensile strength, and VH the Vickers hardness of the material or region in question. $\sigma_{Yield}$ and $\sigma_{Ultimate}$ were determined from the tensile tests previously discussed and averaged $c$ values were used for estimation of the tensile response of the HAZ where direct tensile tests could not be taken due to limited material. The averaged values of $c_{Yield}$ and $c_{Ultimate}$ used for the HAZ were found to be 2.1 and 2.8, respectively.

Specimens for metallurgical analysis were cut through the center of the weld along the rolling direction of the aluminum plates then mounted in acrylic resin before mechanical polishing. Polished specimens were then placed on an optical microscope to measure particle distribution in the three weld regions. After a slight etch using diluted nitric acid, electron backscatter diffraction (EBSD) and SEM were used to determine the grain sizes, orientations, and textures in the separate weld zones. Another specimen was etched with Keller's reagent and viewed under an optical microscope to obtain measurements of the distinct regions of the weld nugget. Finally, in order to obtain an accurate measure of the small precipitates (~10-20 nm diameters) described by Maisonnette et al. [44] and Benedetti et al. [45], samples for high resolution transmission electron microscopy (HRTEM) were ground to 100 μm thickness, reduced to a 3 mm diameter wafer using a specialized punch, dimpled to reduce thickness in the center, and finally thinned to electron transparency by ion milling.

## 3. Results

### 3.1 Tensile and Shear Mechanical Properties

Miniature dog-bone tensile specimens were extracted from the weld section and the parent material to analyze the differences in mechanical strength. Figure 3 shows tensile stress-strain curves produced using these specimens. The error bars represent the standard deviation of the stress results. The error in the parent material varied too little to be noticed in Figure 3 having a maximum standard deviation of approximately 5.4 MPa. The stress-strain curves reveal a significant reduction in tensile strength and ductility for the weld section as compared to the parent material. Fractography reveals large dimples around particles in the parent material suggesting the primary cause of failure to be void growth from large inclusions embedded in the material (Figures. 3a,b). The effect of void growth from these inclusions is much more prominent in the parent material than in the fusion zone, which shows less ductile behavior (nucleation dominant). This suggests that the weld-induced microstructure causes nucleation to be a dominant mechanism in the weld region as compared to void growth in the parent material (see [44–48]).

Figure 3c shows that the failure strains in the parent material reach up to 22.2% while failure strains in the fusion zone reach only 4.8% indicating a 78% drop in failure strain. Decreased ductility is further evidenced by the increase in the strain hardening exponent. The maximum strain hardening exponent increased as a result of welding from 0.163 in the parent material to 0.382 in the fusion zone as determined using the method outlined by Xu et al. [51].

The rolled 6061-T6 aluminum alloy parent material has a yield strength of 312 MPa and an ultimate strength of 341 MPa. In contrast, the welded portion of the material has a yield strength of 120 MPa and an ultimate strength of 162 MPa, i.e., ~62% and 52% lower yield and ultimate strengths, respectively, when compared with the parent material. The tensile results indicate that the microstructural damage accumulation due to the change in number density of voids is much faster in the fusion zone than in the parent material, where the change in void growth is the more dominant and slower process as shown in Figure 3a.

Failure in optimal RSW bonds during lap shear tests occurs in the HAZ as reported by Hayat [52]. While tensile tests could not be performed in the HAZ due to limitations of specimen geometry, it is possible to predict the degree of loss in ductility throughout the weld by estimating the change in strength using the SPT technique. Here, the main objective is to quantify shear properties of the joint that are relatively difficult to measure and are generally absent in the open literature (particularly for RSW joints). Figure 4a shows the shear stress evolution, measured using SPT, as a function of normalized extension for the different sections in the welded region as highlighted in Figure 2c. From the shear punch data shown in Figure 4a the shear yield stress (from a 1% offset criterion) and also the ultimate (maximum) shear stresses could be determined. Figures 4b and 4c show the correlation of measured shear properties with the tensile properties for the parent material and fusion zone. These values are compared to the predicted relation $\sigma_{ys} = 1.77\tau_{ys}$ and $\sigma_{us} = 1.8\tau_{us}$ for the yield and ultimate strength, respectively as mentioned in Guduru et al. [42]. Here, we used the pure aluminum and 6061 data from Guduru et al. [42] for comparison. The shear strength through the weld region was measured along with the ultimate tensile strength and tensile yield strength estimated from the microhardness data (Figure 4d).

Hardness coefficients were calculated in the parent material and fusion zone and the hardness coefficient was assumed to decrease linearly as the weld center was approached. Tensile strength estimates in the HAZ may not be exact as no tensile data was acquired to validate the assumption of a linear decrease in hardness coefficient, but a valid observation can be made regarding the distinct decrease in the material strength of the HAZ. The high temperature experienced in the HAZ reduces the tensile properties such that the shear and tensile properties are nearly equivalent throughout the weld. In the parent material, however, tensile strength is higher than shear strength. The lowest material strength is found near the center of the weld where shear yield strength is about 50% lower than that of the parent material. The loss of the T6 condition, which occurs during welding, is expected to decrease the mechanical strength reflected in the drop of shear properties [44]. Hence, during the weld thermal cycle, the increasing degree of dissolution of the precipitates from the parent material to the center of the weld nugget causes progressive loss of strength, resulting in the observed shear strength profile across the cross section of the weld (Figure. 4a).

To check for anisotropy resulting from the solidification induced microstructure, shear punch tests were performed in the rolling, transverse, and normal directions of the weld and rolled plates. Figure 5 shows the shear stress evolution as a function of normalized extension for the rolling, transverse, and normal directions. Interestingly, Figure 5 shows a minimal dependence on the orientation of the test relative to the original rolling direction of the plate. The parent material and HAZ show increased strength in the normal direction and decreased extension in the rolling direction; however in the fusion zone, anisotropy plays a larger role, which could be attributed to residual stresses, reported in our previous study (Florea et al. [7]), resulting from deformation during loading as well as a sharp gradient in grain misorientation distribution resulting from the intense temperature gradient. Normalized extensions to failure for the transverse and rolling directions are, respectively, 27% and 15% lower than normalized extension to failure of the rolling direction in the fusion zone.

## 3.2 Microstructure Analysis

The variability in the microstructure and mechanical properties developed during the RSW process arise from the integration of mechanical, metallurgical, thermal, and electrical phenomena. Specifically, the interaction between thermal and metallurgical phenomena results in a varied microstructure, i.e., reduced grain size and homogenization of precipitates. In addition, thermal and mechanical phenomena result in non-uniform thermal strains and residual stresses. Electrical and thermal effects strongly correlate through Joule heating, which generates high temperature gradients and, consequently, non-uniform weld strength. From metallurgical and mechanical perspectives, fluctuating mechanical responses among the parent material, HAZ, and FZ involve non-homogeneous distribution of the material microstructure.

### 3.2.1 Grain Orientation Analysis

To explain the detrimental effects of spot welding on the material strength and ductility, various microstructural characterization techniques were performed. EBSD was used to measure grain orientations and texture on a plane transverse to the welding direction, as well as better explain the anisotropic properties observed in Figure 5. Figure 6 shows EBSD results of texture evolution and misorientation distributions in the weld region, which reveal increasing uniformity of crystal orientations as the weld center is approached. There are some minor variations in orientation within the grains of the fusion zone (denoted by the slight color fluctuations across a grain in the inverse pole figures of Figure 6a), which likely result from dendrite formation within these grains. The uniform microstructure of the fusion zone is generated by the high heat of the RSW process leading to morphology of the microstructural features in the regions around the weld.

A small grain size difference occurs as a result of resistance spot welding. Comparing the size of grains found in the fusion zone (Figure 6a) to the grains found in the parent material (Figure 6c), the average grain diameter increases from 21 μm to 24 μm which is not a drastic increase. However, along the rolling direction, the grains decrease by 2% while along the transverse direction, the grains increase by 47% allowing the grains within the fusion zone to exhibit no elongation. Grain size measurements are listed in Table 3. Optical imaging of the etched fusion zone (Figure. 7) reveals large dendrite structures within these grains. The high temperatures achieved in the weld nugget lead to the loss of grain elongation while the thermal gradient between the hot weld center and the cool parent material allows dendrites to form toward the center of the fusion zone.

Further analysis shows that the Taylor factors in each region (Figure 8) indicate resistance to slip along the {111} planes within the fusion zone when loaded in uniaxial tension along the rolling direction of the initial plate, thereby decreasing the ductility of the fusion zone. Interestingly, the Taylor factors in the HAZ show little difference to the Taylor factors of the parent material.

### 3.2.2 Precipitate Analysis

Precipitate strengthened alloys are often composed of precipitate structures at multiple length scales. In the case of this sample of 6061-T6, three primary precipitate sizes are observed. The first of these precipitate structures are the largest particles (diameter > 100 nm) in the matrix. Energy dispersive X-ray analysis (EDX) shows these larger particles to have a high Fe and Cr concentration. These particles weaken the alloy as voids form around these particles under tensile loading as seen from the fracture surface image of Figure 3a. Comparing optical images of the parent material and fusion zone, Figures 9a and 9b respectively, the high heat of the weld process is seen to decrease the size of these particles (approximately a 13% decrease from the parent material to the fusion zone) causing them to nucleate more voids as seen in the fracture surface image of Figure 3b.

The second precipitate is a Cu rich precipitate that has an intermediate size (<100 nm). This precipitate structure was not investigated further, because it was not observed on fracture surfaces, and it is deemed too large to noticeably contribute to precipitate strengthening in comparison to the other precipitates. The final precipitate structure is the smallest precipitate and is mainly composed of Mg and Si. In the parent material, these precipitates are small (approximately 20-50 nm length) needle-shaped precipitates found by HRTEM combined with EDS (Figure 10) and contribute to strengthening of the 6061-T6 alloy by pinning dislocations. As dislocations are slow to shear through or loop around these precipitates, the material exhibits higher yield and ultimate strengths. As the 6061-T6 alloy is heated during welding, these precipitates agglomerate as observed by Maissonette et al. [44] where it was found that high temperatures contribute to fewer precipitates in the material. Further, Figure 11 shows dislocations bowing around precipitates, which is in agreement with the previous work of Dirras et al. [53].

Traditionally, 6061 is a precipitate strengthened material implying that the loss of stress from the welding process should primarily be a result of a loss of the T6 condition with minimal grain size strengthening effects. For the samples investigated here with only a 3 μm increase in grain size resulting from the welding process, the Hall-Petch relationship, with $\sigma_0 = 50$ MPa and $k_y = 0.326$ MPa m$^{1/2}$ [54], only accounts for a decrease in yield stress of less than 5 MPa. This indicates that the nearly 200 MPa drop in yield stress observed in the fusion zone is a result of the loss of the T6 condition in the welded region.

In summary, this report provides a more complete understanding of the correlation between the microstructural change and the decrease of mechanical strength and ductility resulting from a RSW process on 6061-T6 aluminum alloy. The primary strengthening mechanisms and the cause for the loss of ductility are identified in the RSW 6061-T6 alloy. The understanding obtained through this research will allow enhanced design of joints in aluminum structures leading to the development of lighter weight technologies.

## 4. Conclusions

The structure-property relationships of RSW 6061-T6 aluminum alloy lap joints were characterized using tensile, shear punch, and microhardness tests coupled with optical, scanning, and transmission electron microscopy. The welding process was shown to reduce the strength as well as the ductility of the weld. The decreased mechanical properties were linked to residual stresses and metallurgical phenomena including agglomeration of precipitates and decreased grain and particle sizes.

Resistance spot welding is shown to reduce the tensile and shear mechanical response of 6061-T6 aluminum alloy. The decrease in mechanical strength is coupled with a decrease in the maximum strain hardening exponent which indicates decreased ductility in the fusion zone as a result of RSW. The decreased mechanical response is a result of the changes in the microstructure resulting from RSW. The microstructural changes have a profound effect on the shear properties, but a lesser effect on the tensile properties likely resulting from residual stresses reported by Florea et al. [7] which cause anisotropy in the fusion zone.

The lowered mechanical properties of the welded material are the result of a combination of microstructural changes such as large Fe-Si particle refinement, fine Mg-Si precipitate accumulation, and a development of more uniform texture resulting in an increased Taylor factor. The refinement of large Fe-Si particles is found to decrease the ductility of the welded region through the nucleation of more voids under tensile load than found within the parent material as observed through facture surface imaging. EBSD images show more uniform grain orientations within the fusion zone which contribute to an increased Taylor factor. An increase in Taylor factor within the fusion zone indicates a lower propensity for slip as a result of tensile loading along the rolling direction. This may contribute to the lowered ductility observed within the fusion zone. Little to no change in Taylor factor is observed between the parent material and HAZ. As a result of high temperatures during the welding process, fine Mg-Si precipitates accumulate in the fusion zone reducing the precipitation strengthening observed in 6061 in the T6 condition. The combined effects of precipitate accumulation with slight grain growth within the fusion zone work together to reduce the mechanical strength of the welded joint.

In summary, the reduction in mechanical properties resulting from resistance spot welding in 6061-T6 aluminum alloy is a function of residual stress, particulate size reduction, grain orientation alignment, and precipitate growth. Precipitate agglomeration in the fusion zone likely has the greatest negative influence on the mechanical strength of the weld and finer particle distribution is likely to have the greatest influence on the ductility of the weld. Finally, to our knowledge, this is the first study where shear properties across RSW 6061-T6 are characterized using SPT. Quantifying the influence of process induced microstructural changes (structure) and subsequent effects on mechanical properties is technologically important in advanced energy, transportation, and manufacturing industries.

## Acknowledgements

The authors would like to thank Abdullah Sherif and the Center for Advanced Vehicular Systems at Mississippi State University, the LeRoy Eyring Center for Solid State Science at Arizona State University, and the Army Research Laboratory for their help and the use of their equipment throughout this investigation. Special thanks to the late Dr. Radu Florea for graciously sharing his preliminary work.

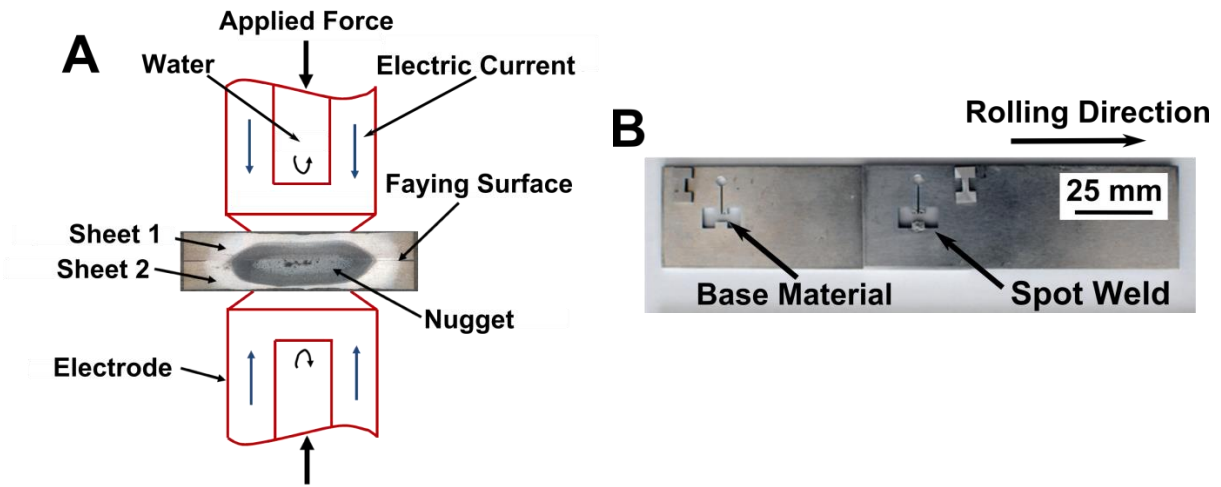

Fig. 1. a) An illustration of the RSW process, b) a RSW-produced 6061-T651 lap joint with tensile specimens cut by wire electric discharge machining (EDM)

**Table 1**
**Welding process parameters**

| | |
|---|---|
| Electrode Force | 3.8 kN |
| Time | 0.115 s |
| Current | 30 kA |
| Voltage | 8 VDC |
| Water Flow Rate | 4 liters/min |

**Table 2**
**Tensile Specimen Dimensions**

| | |
|---|---|
| Gauge Length | 4.75 mm |
| Gauge Width | 2 mm |
| Gauge Thickness | 2 mm |

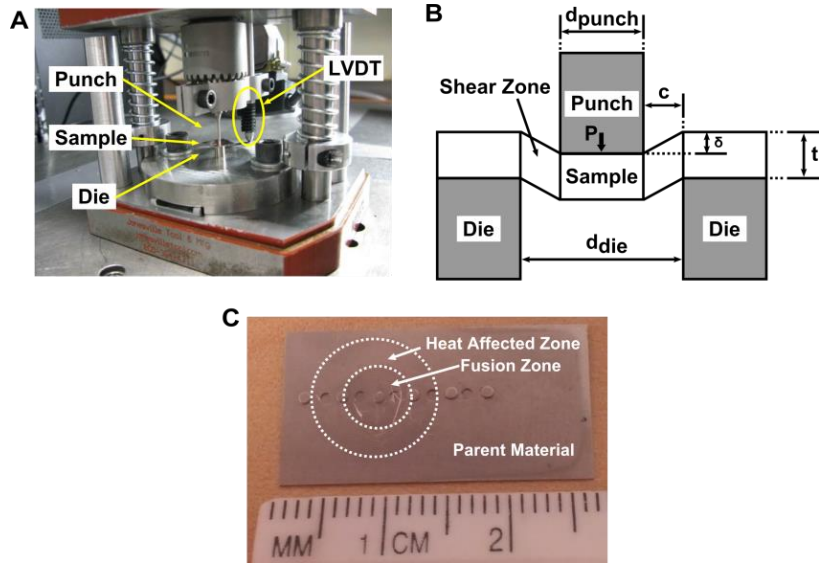

Figure. 2. a) Shear punch setup, b) cross-sectional schematic of shear punch test, and c) location of shear punch testing with respect to weld zone.

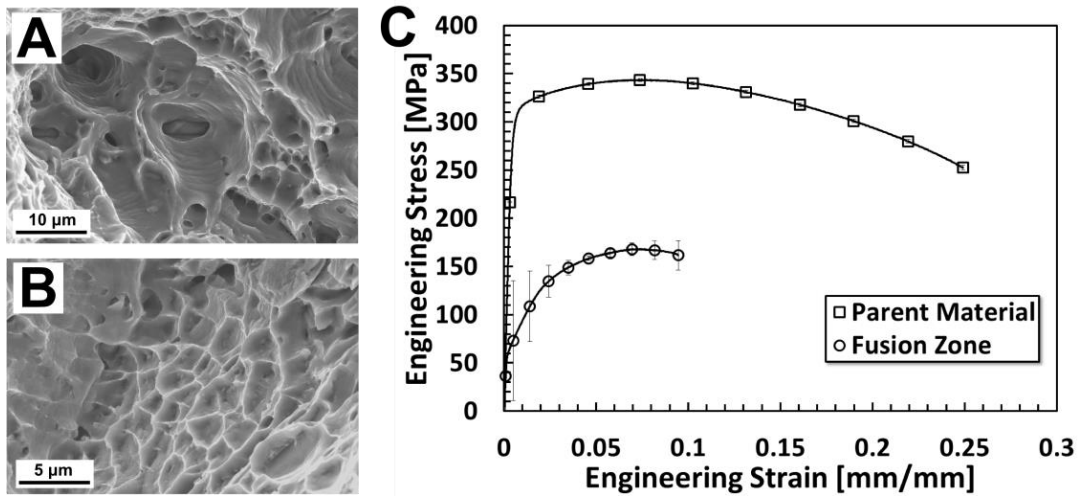

Figure 3. Fracture surface images taken from the a) parent material and b) fusion zone after tensile failure at a strain rate of $10^{-3}$ $s^{-1}$. c) Corresponding tensile stress-strain results taken along the rolling direction. The large dimpled surface of the parent material, indicating high ductility, is caused by void nucleation, growth, and coalescence from large particles within the material.

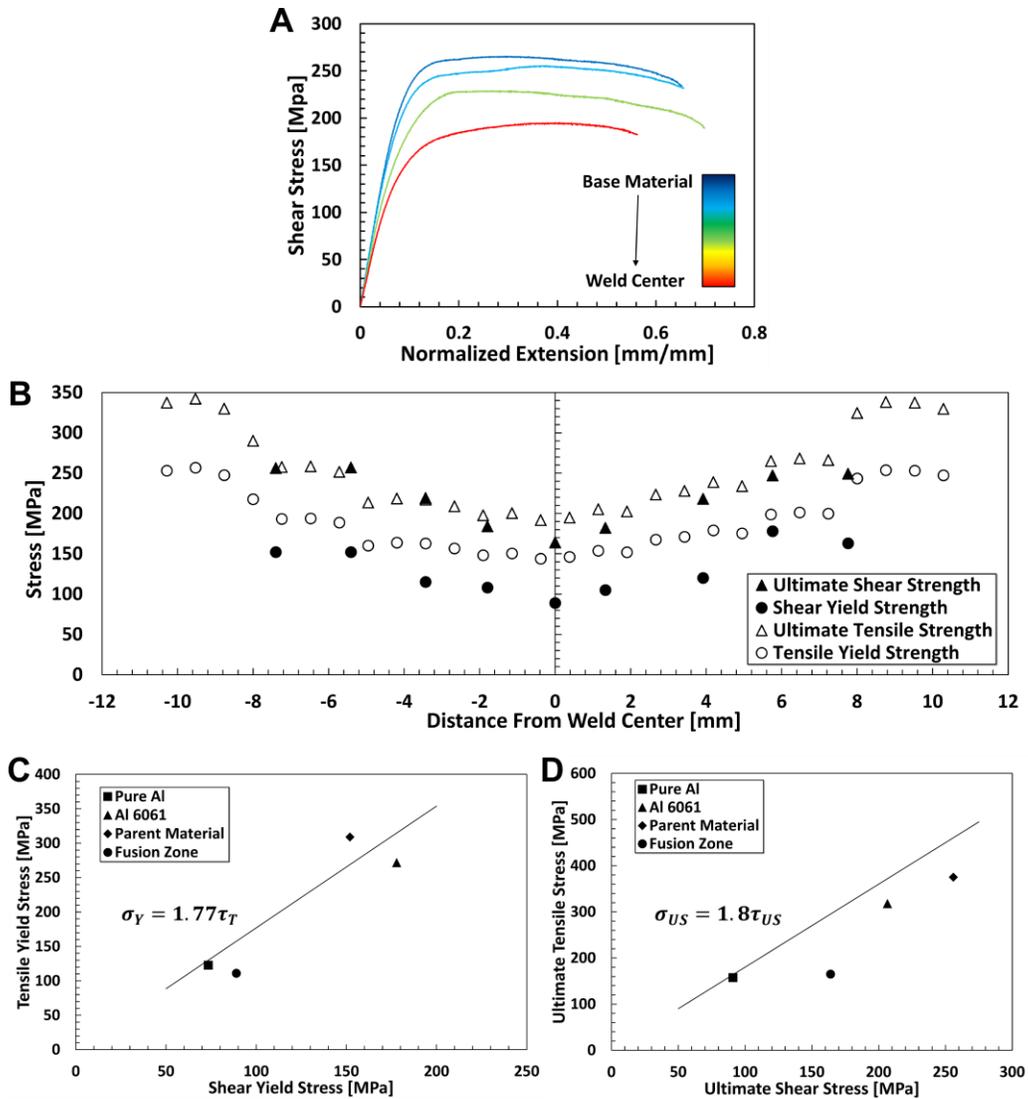

Figure 4. a) Results from shear punch testing through the spot weld. b) Shear yield strength and ultimate shear strength from shear punch testing as well as tensile yield strength and ultimate tensile strength as determined by Vickers hardness values all plotted as functions of position through the weld. The deviations of tensile and shear stress behavior from experimentally determined fits for the c) yield and d) ultimate stress values. Pure Al and 6061 values come from Guduru et al. [38].

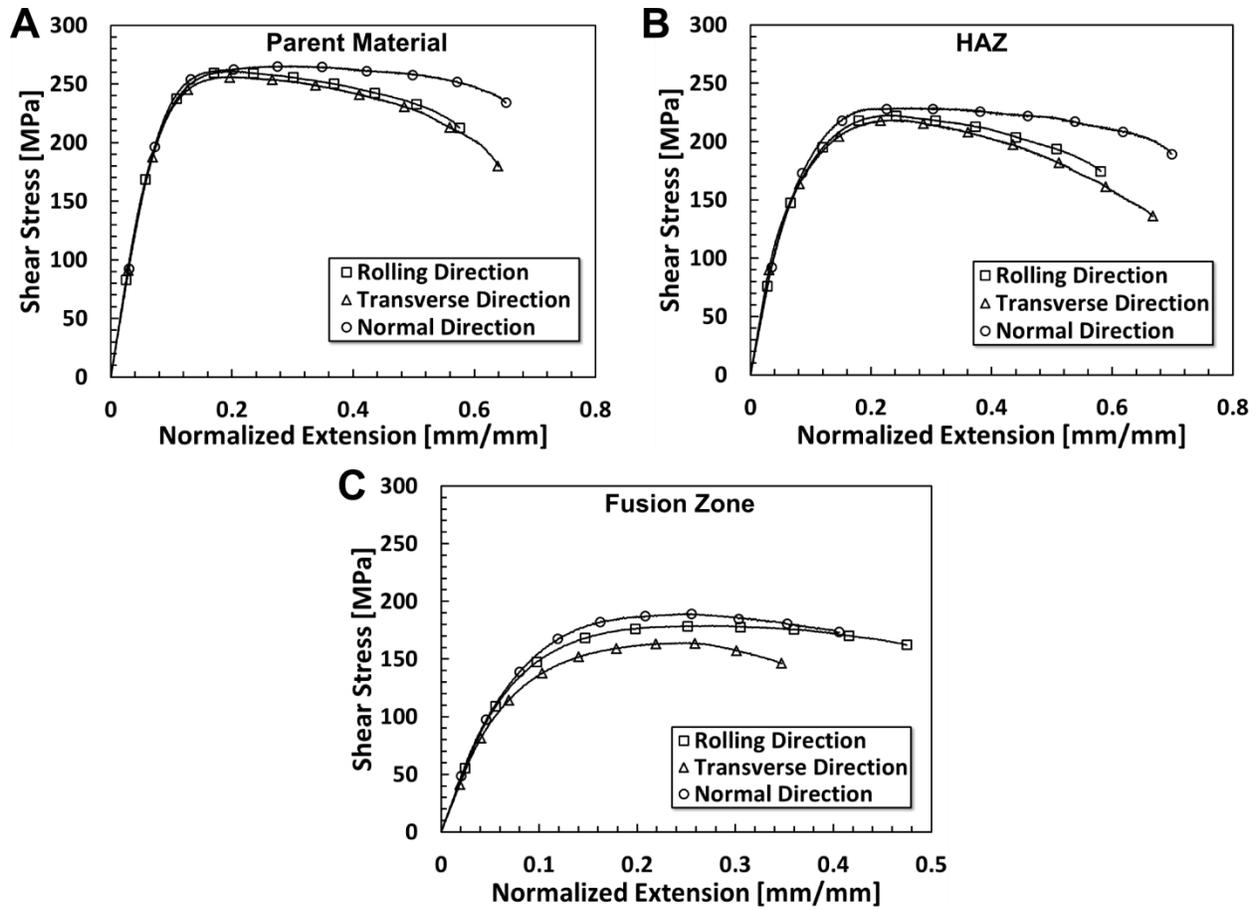

Figure 5. Engineering shear stress responses as a function of the normalized extension along the rolling, transverse, and normal directions for the a) parent material, b) HAZ, and c) fusion zone. The fusion zone shows increased mechanical response dependence on directionality as noted by the fluctuation in the stress values in (c) when compared to those in (a) and (b).

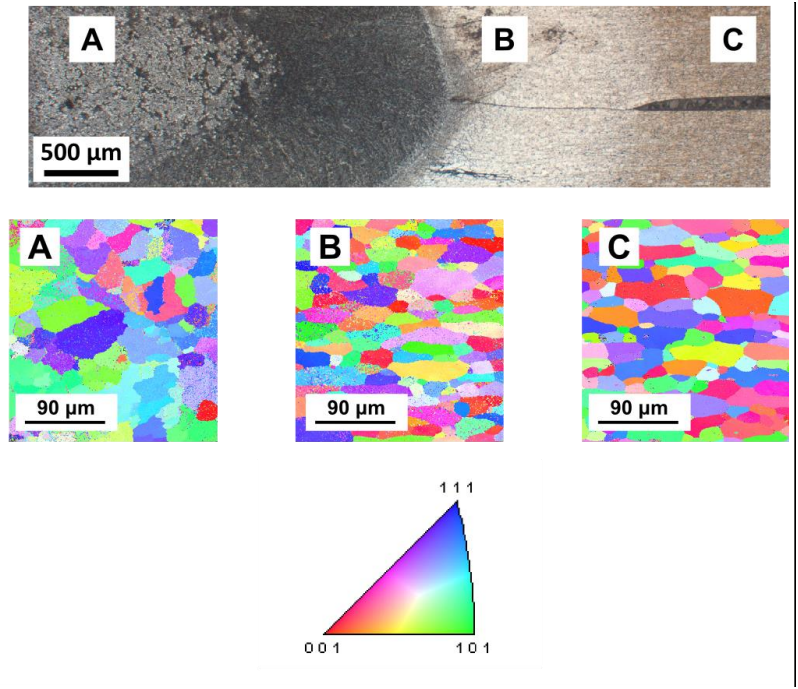

Figure 6. Thermomechanical welding operations often involve high strains and deformation temperatures that result in microstructures that continually evolve away from the parent of the material as a result of significantly altering the material behavior and geometrical dimensions. Inverse pole figures show grains colored by orientation in the a) fusion zone, b) HAZ, and c) parent material. The misorientation angles are based on the rolling direction of the plate. Elongated grains from the rolling process used to form the parent material are seen in both the HAZ and parent material. No elongation is seen within the center of the weld nugget.

**Table 3** Average grain diameters in each of the three regions of the weld.

| Average Grain Diameter [μm] | Parent | HAZ | Fusion Zone |
|---|---|---|---|
| All Directions | 21 (+/- 11 μm) | 19 (+/- 6 μm) | 24 (+/- 6 μm) |
| Normal Direction | 15 (+/- 0.4 μm) | 14 (+/- 1 μm) | 22 (+/- 1 μm) |
| Rolling Direction | 28 (+/- 5 μm) | 24 (+/- 0.4 μm) | 27 (+/- 3 μm) |

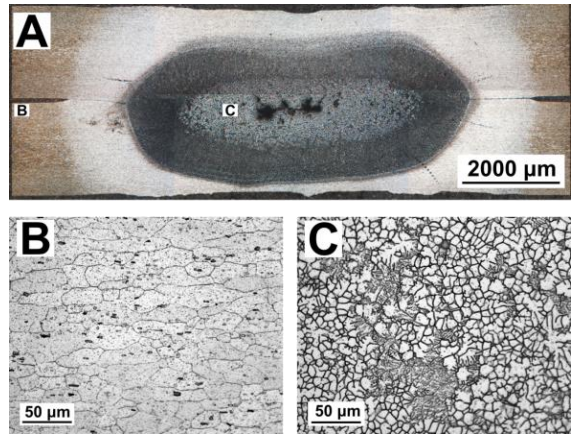

Figure 7. Optical micrographs reveal a) the regions of the weld nugget, b) grains in the parent material, and c) dendrite formation in the fusion zone.

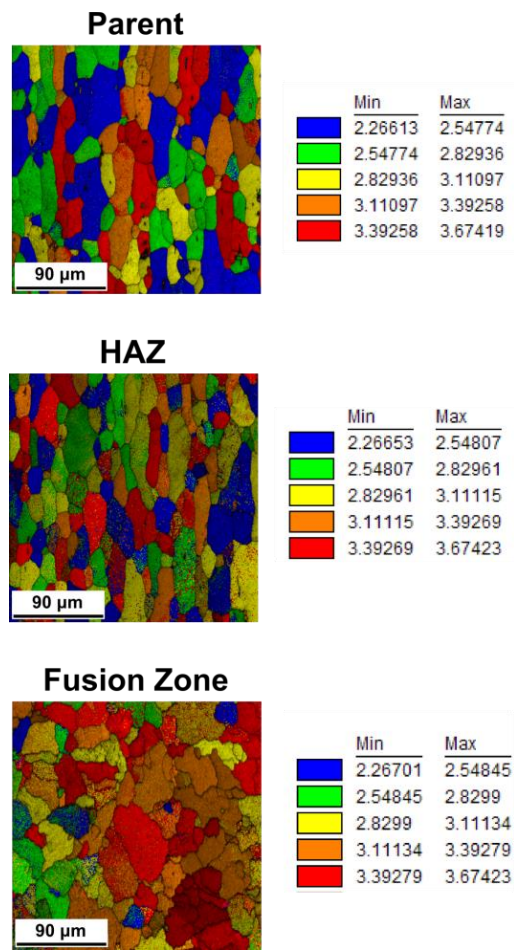

Figure 8. Taylor factor maps indicating a rise in Taylor factor within the fusion zone. The color scale relating to the binning of Taylor factor values is found to the right of each image. Little

change in Taylor factor is noted in the HAZ suggesting that the mechanical strength in the HAZ is not greatly affected by a change in the orientation of the grains.

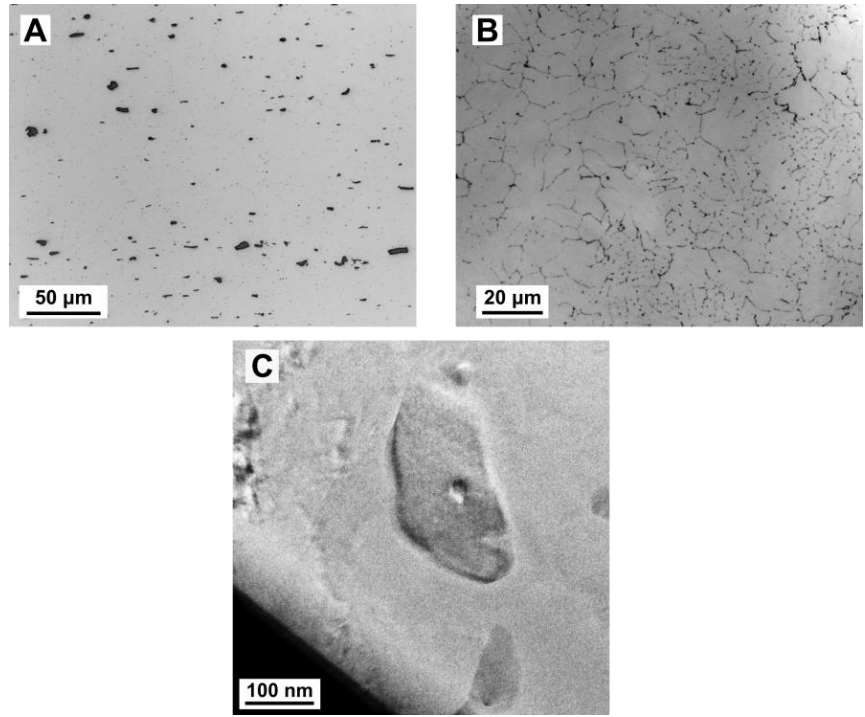

Figure 9. Optical micrographs revealing particles in the a) parent material and b) fusion zone, and c) a TEM image of a particle composed of Fe and Cr in the fusion zone. The RSW process reduced the size of the large Fe-Cr particles in the fusion zone (on the order of a few hundred nanometers), which led to a less ductile material response.

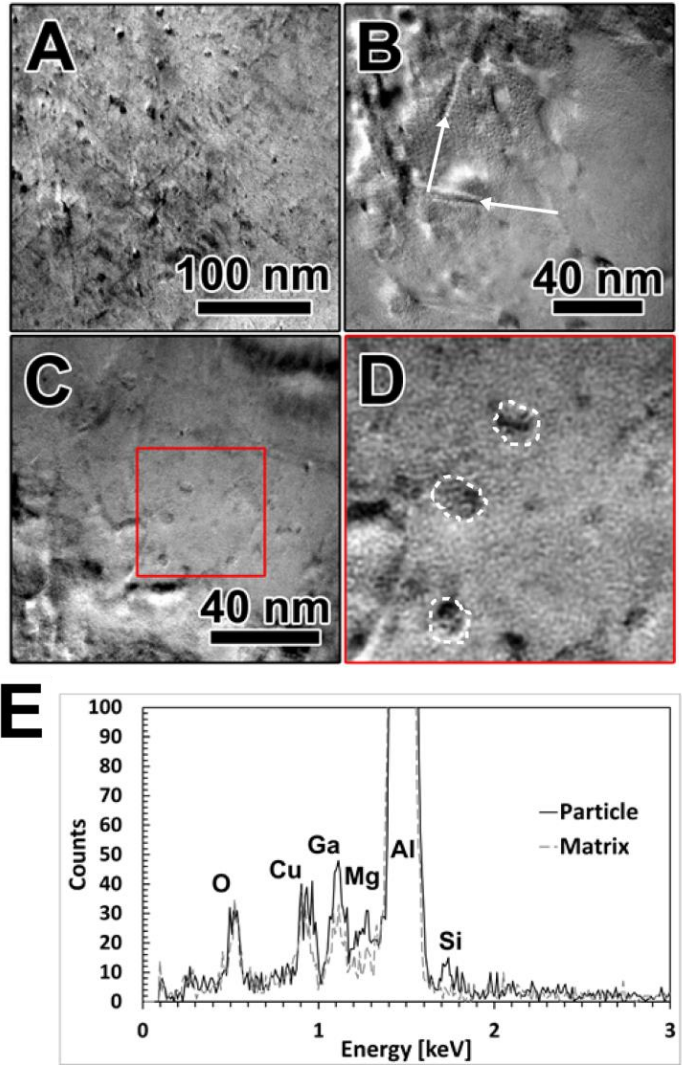

Figure 10. a-b) Precipitates in the parent material are needle shaped (highlighted with arrows). A diffraction pattern over a precipitate (b-top left) shows little to no difference from the diffraction pattern taken over the surrounding material (b-top right) as the precipitates are too small to be individually determined. c-d) No needle shaped precipitates are evident in the fusion zone, but a few spheroidal precipitates (outlined in the magnified region d) are noticed. e) EDS spectra show that the precipitates are mainly composed of Mg and Si. The Ga peak comes from residual Ga ions from sample preparation using a focused ion beam (FIB) and the Cu peak comes from the holder in the microscope. The Mg and Si peaks are small due to the small size of the precipitate compared to the spot size of the EDS.

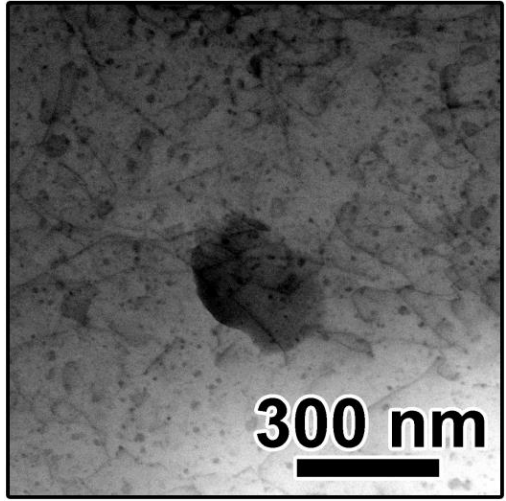

Figure 11. Dislocations are seen bowing around precipitates, but no Orowan loop formation is observed.